# Methods – Pressure control apparatus for lithium metal battery

Bingyu Lu[1], Wurigumula Bao[1], Weiliang Yao[2], Jean-Marie Doux[1], Chengcheng Fang[3*], Ying Shirley Meng[1,4*]

[1]Department of NanoEngineering, University of California San Diego, La Jolla, CA 92093, USA.

[2]Materials Science and Engineering, University of California San Diego, La Jolla, CA 92093, USA

[3]Department of Chemical Engineering and Materials Science, Michigan State University, East Lansing, MI 48824, USA.

[4]Pritzker School of Molecular Engineering, University of Chicago, Chicago, IL 60637, USA.

[*]Correspondence to: shirleymeng@uchicago.edu (Ying Shirley Meng), cfang@msu.edu (Chengcheng Fang)

**Abstract:** Lithium (Li) metal anodes are essential for developing next-generation high-energy-density batteries. However, Li dendrite/whisker formation caused short-circuiting issue and short cycle life have prevented lithium metal from being viably used in rechargeable batteries. Numerous works have been done to study how to regulate the Li growth in electrochemical cycling by using external stacking forces. While it is widely agreed that stack pressure positively affects the lithium plating/stripping process, the optimized pressure range provided by different works varies greatly because of the difference in the pressure control setup. In this work, a pressure control apparatus is designed for Li metal batteries with liquid and solid-state electrolytes (SSE). With considerations of minimizing cell to cell variation, a reusable split cell and pressure load cell are made for testing electrochemical cells with high precision pressure control. The capability of the designed setup is demonstrated by studying the pressure effect on the Li plating/stripping process.

**Introduction:**

The world has witnessed tremendous development in portable electronic devices and electric vehicle technology over the last four decades..[1,2] With its high theoretical capacity (3,860 mAh g$^{-1}$, or 2,061 mAh cm$^{-3}$) and low electrochemical potential (–3.04 V versus the standard hydrogen electrode), lithium (Li) metal has been regarded as the "Holy Grail" anode material candidate for next-generation batteries.[3,4] Still, many formidable obstacles need to be overcome before Li metal could be used practically as the anode in commercial cells. The fundamental problems of using a Li metal include the continuous formation of 1) Solid Electrolyte Interface (SEI), 2) Li dendrites/whiskers and 3) "inactive" or "dead" Li.[5] Because of the highly negative electrochemical potential of Li$^+$/Li, electrolyte components can easily decompose and form SEI on the surface of Li metal. However, the fragile structure of SEI could easily be fractured and reformed during the plating and stripping process, leading to the heterogeneous ionic conductivity of the SEI. The cracks on the SEI and the inhomogeneous ionic conductivity together cause the growth of Li dendrite[6]. In the stripping process, the high tortuosity of the Li dendrite could easily isolate metallic Li from electron-conducting materials, causing so-called "dead" Li. The formation of the "dead" Li during the cycling eventually leads to the low Coulombic efficiency (CE) and safety hazards of the Li metal batteries.

Recently, studies have been focusing on using external mechanical force to control the growth of Li morphology in both liquid electrolytes and SSE.[7–15] The effect of external stacking pressure was first studied by Dr. Wilkinson and his colleagues in 1991.[16] By designing an *in situ* cell, they could monitor the volume expansion and pressure evolution of the Li metal battery while it is cycling. It was discovered that mechanical pressure greatly affects the morphology of the electrochemically deposited Li (EDLi). The results indicate that at low stacking pressure (below 140 kPa), the EDLi was in a needle-like structure, while at large stacking pressure (1400 kPa), the EDLi formed a column-like structure, which was dense and uniform. They believed the improved morphology was the reason why the CE of the cell with high stacking pressure was better. Inspired by this conclusion, many recent works have studied the importance of pressure on Li metal anode. Anode-free pouch cells with external stacking plates were mostly used to study how various pressure (0 to 2000 kPa) could affect the cyclability of the cell.[17–20] The results show that as the pressure increased, the capacity retention of the anode-free cell would generally improve. However, the amount of improvement was related to the type of cell setup, and sometimes high pressure would even hinder the cell's performance. The precise control of stack pressure on the all-solid-state battery (ASSB) is also crucial for cycling stability. Internal shorting caused by Li dendrite formation is prevalent during ASSB cycling. Porz *et al.* claimed that Li could penetrate even near-perfectly dense single crystal oxide solid electrolytes (such as $Li_7La_3Zr_2O_{12}$) during the cycling, leading to the shorting of the cell as a result.[21] Doux *et al.* further explored the effects of the stack pressure on the cycling stability of the ASSB. It was found out that the optimized stack pressure of 5 MPa could enable Li metal anode to cycle at room temperature without shorting for more than 1000 hours.[15,22]

As numerous studies have been done trying to determine the optimized pressure ranging for Li metal anode cycling, it has been brought to attention that a carefully designed pressure control setup is needed to quantitatively study the effects of pressure on the cycling behavior of lithium metal anode. In this work, a split cell coupled with a pressure load cell (a pressure sensor) is designed for precisely controlling and monitoring the stacking pressure of Li metal cells, with a resolution as high as 0.1 kPa. The effect of pressure on the Li plating/stripping process is also demonstrated as an example.

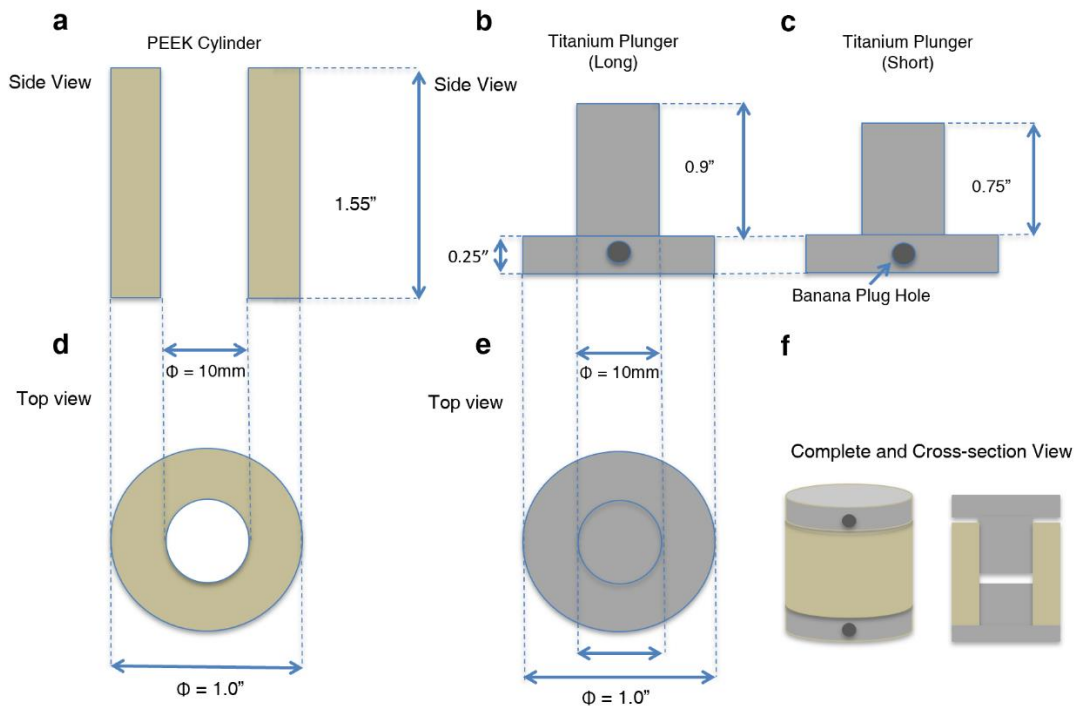

Figure 1. Dimensions for the split cell setup. Side views of a) PEEK Cylinder; b) Long Titanium Plunger; c) Short Titanium Plunger. Top view of d) PEEK Cylinder; e) Titanium Plungers. f) Complete and cross-sectional view of the split cell setup.

## Pressure Control Apparatus

### Split cell:

The split cell consists of two parts (Fig. 1): one polyether ether ketone (PEEK) die (Fig. 1a, d) mold and two titanium (Ti) plungers (Fig. 1b, c, e). The Ti plungers have dual purposes in the split cells: 1) to work as current collectors and 2) to apply pressure. Therefore, Ti is chosen as the material for the plungers because of its high electronic conductivity, high mechanical strength, and high stability towards chemical corrosion. One hole is drilled on the bottom of the plungers to provide space for the banana plugs that are connected to the battery cyclers. The PEEK die mold is used as a container to hold the electrochemical cell in place. The total length of the Ti plungers (1.65 inches) is slightly higher than the length of the PEEK die (1.55 inches) so that the PEEK die can freely adjust its position and will not affect the pressure control. All parts required high precision machine polishing so that the friction between the Ti plunger and the inner wall of the PEEK die mold is at a minimum. It is important that the friction between the titanium plungers and the PEEK die does not influence the pressure control while providing a good sealing for the electrochemical cell inside (Fig. 1f).

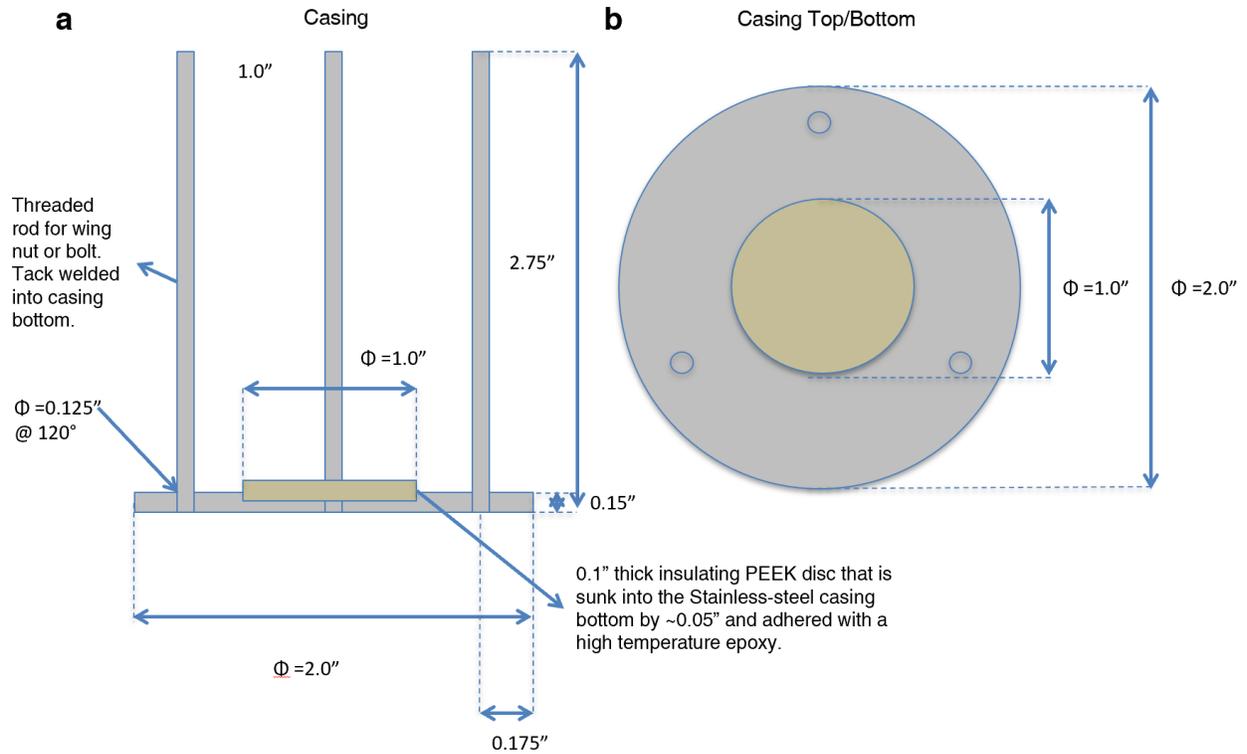

Figure 2. Dimensions for the casing for the split cell and load cell. a) Side view of the casing rods and bottom; b) Top view of the Top and Bottom of the casing.

**Casing:**

The casing consists of two parts (Fig. 2): three threaded rods (Fig. 2a) and top/bottom plates (Fig. 2b). The split cell and the pressure load cell are placed uniaxially inside the casing (Fig. 3a). The top/bottom plates stack the split cell and the pressure load cell. Bolts or wing nuts are used to precisely control the stack pressure exerted by the plates. A thin PEEK disc is sunk into the top/bottom plates at the center where the split cell will sit. The PEEK disc can prevent shorting. All the metal parts of the casing are made of stainless steel.

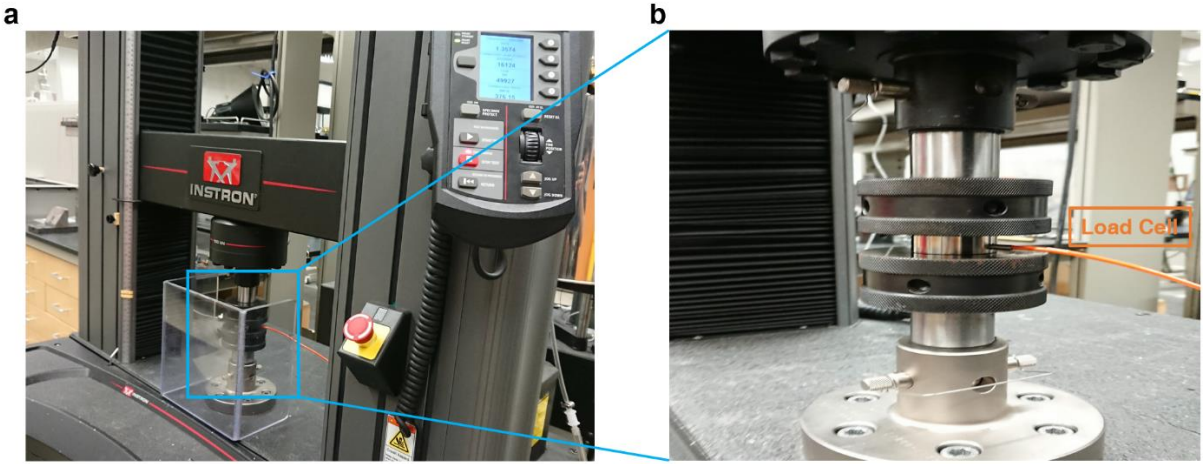

Figure 3. a) the calibration of the load cell using Instron 5982 Universal Testing System. b) Zoom-in image of the load cell.

**Load cell calibration:**

The DYHW-116 load cell (Bengbu Dayang Sensing System Engineering Co, Ltd) and a digital LED display monitor the exerted stacking pressure on the split cell. The load cells have different measuring ranges. Choosing the most suitable range (0 to 1000 kPa is enough for Li metal anode with liquid electrolyte) for the specific testing scenario is beneficial for a higher pressure resolution. Besides, it is crucial to calibrate the load cell before use since the default reading of the load cell might be way off from the actual pressure that is being measured. To calibrate the load cell, a set of known loads is first applied onto the load cell by using the 100 kN Instron 5982 Universal Testing System (Figure 3a). The readings from the load cell are first calibrated at the end of the range of the load cell (i.e., 1000 kPa for the load cell described earlier). Then, the accuracy of the load cell setup can be compared to the load applied by the Instron testing system over the whole range of the load cell to ensure that the calibration is accurate. This calibration method allows to precisely set the load cell readings to correspond to the actual pressure applied to the sample by taking into account the sample area.

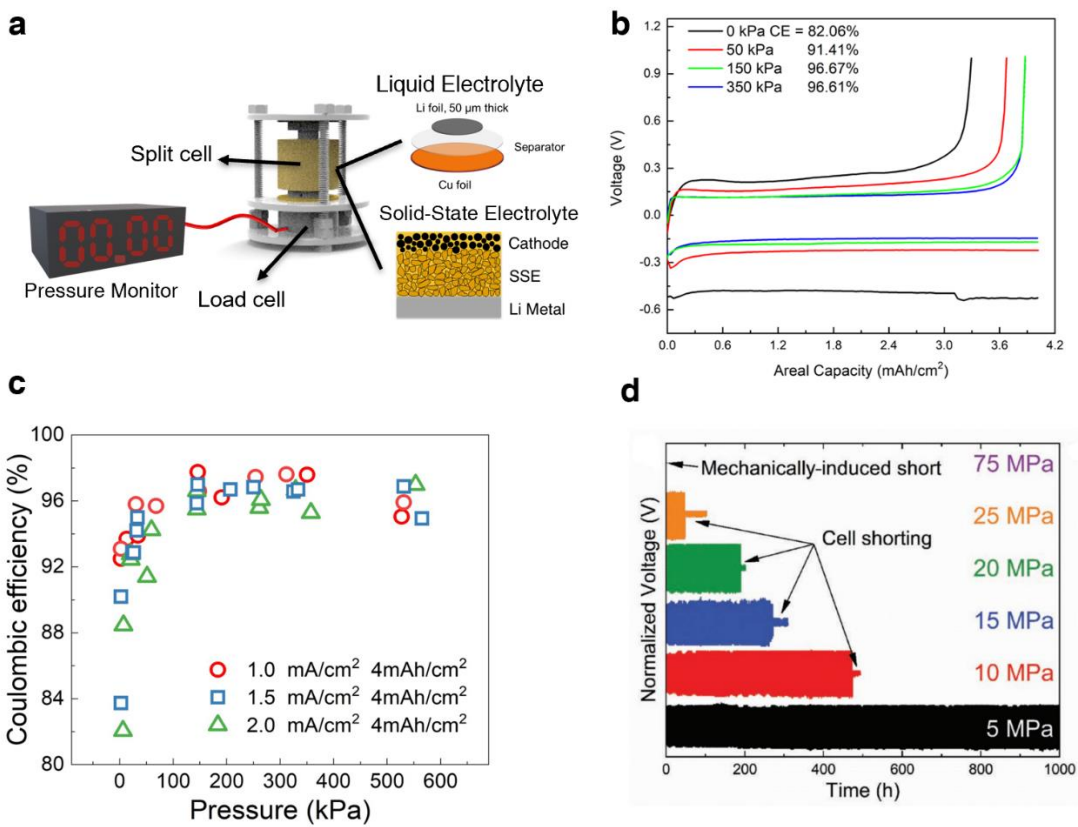

Figure 4. a) Completed pressure control apparatus for lithium metal battery with liquid electrolyte or SSE. b) The voltage profile of first cycle plating/stripping test of the Li||Cu cells with current a density of 2 mA cm$^{-2}$ in high concentration ether-based electrolyte: 4.6 m LiFSI + 2.3 m LiTFSI in DME. c) The trend of first cycle CE under different stack pressures, at current densities of 1.0, 1.5, and 2.0 mA cm$^{-2}$, all plated for 4 mAh cm$^{-2}$ and stripped to 1 V in a liquid cell.[11] d) Normalized voltage profiles of Li symmetric cells as a function of time during plating/stripping at different stack pressures in the solid-state cell.[15] Reprint permission obtained.

**Example: pressure effects on the Li||Cu plating/stripping process**

With liquid electrolyte: The Cu||Li cell is assembled in an Ar-filled glovebox. The Cu||Li cell is made by sandwiching a thin Li metal foil (7 mm diameter, 50µm thick), Celgard 2325 separator (10 mm diameter), and the cleaned Cu foil (10 mm diameter) between the two Ti plungers inside the PEEK die mold. It is recommended to use thin Li metal foil (less than 50 µm) in the split cell because thicker lithium foil (higher than 500 µm) will deform significantly under high pressure, leading to a relaxation of pressure during testing or even short-circuiting. Only a minimum amount of electrolyte (~5 µL) is required for the Cu||Li cells. After the assembly, the split cell onto the load cell and then fixed by the cell casing, which provides the uniaxial stacking pressure (Fig. 4a). The three nuts adjust the uniaxial stacking pressure on the cell casing. The nuts need to be carefully adjusted in order to apply the desired stacking pressure. It is important

to adjust the split cell in a vertical position so that the applied pressure can be uniformly distributed.

Fig. 4b demonstrates the first cycle charge-discharge profile of Li||Cu cell under different pressure with liquid electrolyte. The CE is directly related to the external pressure exerted on the cell. At 0 kPa (no external pressure, only gravitational force), the plating process shows the highest overpotential (-0.48 V), and the CE is only 82.06% even with a high concentration ether-based electrolyte.[11,23] With a slight increase of the stacking pressure (50 kPa), the plating overpotential decreased to -0.22V, and the CE increased to 91.41%. With the further increase of the exerted pressure, the CE stabilized at around 96.6%, confirming that the pressure can effectively facilitate the plating and stripping process of the lithium metal. The trend of first cycle CE under different stack pressures, at a range of current densities (1.0, 1.5, and 2.0 mA cm$^{-2}$) is also shown in Fig. 4c. It is clear that the stack pressure significantly affects the plating/stripping CE of Li||Cu cells with liquid electrolytes. It should be noted that such split cell is an open cell design, that may bring concerns about the cycle life in this setup since the amount of liquid electrolyte is limited.

With solid state electrolyte: Fig. 4d further illustrates the pressure effect on the stability of Li plating/stripping with an argyrodite SSE. Li symmetric cells with $Li_6PS_5Cl$ are cycled at 75 µA cm$^{-2}$, with 1 h plating/stripping cycles until short-circuiting is observed. The cell pressed to a stack pressure of 75 MPa was short-circuited before the test began. Because of the low yield strength of Li metal, the creeping of Li under such a high pressure allows it to flow within the pores of the SSE, creating an electronic pathway that shorts the cell internally even before any electrochemical testing. When a stack pressure of 25 MPa is applied, the symmetric cell can be cycled for around 48 h before short-circuiting is observed, which is indicated by a sudden drop in the overpotential. It is noteworthy that the cells with stack pressure of 25 MPa and lower only shorts during the plating and stripping process. Similar cycling tests were conducted at stack pressures of 20, 15, and 10 MPa, and similar shorting behavior was observed after 190, 272, and 474 h, respectively. However, at a stack pressure of 5 MPa, no short circuit was observed within 1000 h of the plating and stripping test. All these results show that Li symmetric cell short-circuiting is both a mechanical and an electrochemical phenomenon. A clear trend can be observed between stack pressure and the time needed before short-circuiting occurs.

**Conclusion**

In summary, a pressure control apparatus is designed for lithium metal batteries with liquid or solid electrolytes. The pressure control apparatus consists of one split cell, one load cell, and a metal casing. With high precision machine polishing, the split cell can host liquid electrolytes inside its PEEK die and provide stacking pressure to the electrochemical cell. After careful calibration, the pressure load cell can in situ monitor the pressure exerted on the electrochemical cell with a resolution as high as 0.1 kPa. The apparatus is then used to study the pressure effects on the plating/stripping process in both liquid and solid-state cells. It is shown that the stability of Li plating/stripping is directly related to the external pressure exerted on the cell. This work provides guidelines for designing a pressure control setup for Li metal batteries.

Acknowledgment

Funding: This work was supported by the Office of Vehicle Technologies of the U.S. Department of Energy through the Advanced Battery Materials Research (BMR) Program (Battery500 Consortium) under Contract DE-EE0007764.